\documentclass[aps,prd,reprint,superscriptaddress]{revtex4-1}

\usepackage[english]{babel} 
\usepackage[T1]{fontenc}    

\usepackage{amsmath,amssymb,bm}

\usepackage{graphicx}

\usepackage[pdftex]{hyperref}       
\hypersetup{colorlinks,             
            linkcolor=blue,        
            anchorcolor=black,      
            citecolor=blue,        
            filecolor=black,        
            menucolor=black,        
            urlcolor=blue,         
	        bookmarksopen=true,     
	        bookmarksnumbered=true, 
	        pdftitle={Estimating the thermally induced acceleration of the New Horizons spacecraft},
            pdfauthor={Andre Guerra, Frederico Francisco, Paulo Gil, Orfeu Bertolami},
            pdfsubject={Article estimating the non-gravitational acceleration of the New Horizons spacecraft},
            pdfstartview=FitH,
            pdfdisplaydoctitle=true,
			pdfpagemode=UseNone} 

\newcommand{\unit}[1]{\ensuremath{\mathrm{#1}}}		

\begin{document}


\title{
Estimating the thermally induced acceleration of the New Horizons spacecraft
}  

\author{
Andr\'{e} G. C. Guerra
} 
\email{aguerra@fc.up.pt} 
\thanks{Corresponding author}
\affiliation{Departamento de F\'{i}sica e Astronomia and Centro de F\'{i}sica do Porto, Faculdade de Ci\^{e}ncias, Universidade do Porto, Portugal}  

\author{
Frederico Francisco
} 
\email{frederico.francisco@fc.up.pt} 
\affiliation{Departamento de F\'{i}sica e Astronomia and Centro de F\'{i}sica do Porto, Faculdade de Ci\^{e}ncias, Universidade do Porto, Portugal}  

\author{
Paulo J. S. Gil
} 
\email{paulo.gil@tecnico.ulisboa.pt} 
\affiliation{CCTAE, IDMEC, Instituto Superior T\'{e}cnico, Universidade de Lisboa, Av.\ Rovisco Pais, 1049-001 Lisboa, Portugal}  

\author{
Orfeu Bertolami
} 
\email{orfeu.bertolami@fc.up.pt} 
\affiliation{Departamento de F\'{i}sica e Astronomia and Centro de F\'{i}sica do Porto, Faculdade de Ci\^{e}ncias, Universidade do Porto, Portugal}  

\date{\today}

\begin{abstract}
    Residual accelerations due to thermal effects are estimated through a model of the New Horizons spacecraft and a Monte Carlo simulation. We also discuss and estimate the thermal effects on the attitude of the spacecraft. The work is based on a method previously used for the Pioneer and Cassini probes, which solve the Pioneer anomaly problem. The results indicate that after the encounter with Pluto there is a residual acceleration of the order of $10^{-9}~\unit{m/s^2}$, and that rotational effects should be difficult, although not impossible, to detect.
\end{abstract}


\maketitle                


\section{Introduction}
\label{sec:Introduction}

The first probes aimed at the planets beyond the asteroid belt were launched in the 1970s, starting with the appropriately named Pioneer 10 and 11. These two spacecraft visited Jupiter and Saturn and paved the way for the two much heavier and more sophisticated Voyager 1 and 2, which completed the round through all four gas giants in the Solar System. More recently, orbiter missions like Galileo, launched in 1989, and Cassini, launched in 1997, have been sent to explore Jupiter and Saturn, respectively, and some of their moons.

The now demoted dwarf planet Pluto had yet to be visited by a man-made probe. New Horizons was the first spacecraft to explore Pluto and its moon Charon. Launched in January 2006, its closest approach to Pluto was on July 2015, after a nine and a half year trip. The main objective was to perform a series of scientific studies of what was, at the time, still classified as a planet~\cite{Fountain2008}.

A common feature of all of these probes is their power source. Since solar panels are unable to generate the required amount of energy beyond the asteroid belt, these spacecraft are powered by Radioisotope Thermal Generators (RTGs). These devices generate large quantities of excess heat, since the power conversion through the thermocouples has a rather low efficiency.

In fact, the controversy about the so-called Pioneer anomaly, which lingered around for a decade~\cite{Anderson1998,Anderson2002}, led to the discovery that the anisotropic heat radiation from these spacecraft had indeed a detectable impact on the spacecraft's trajectory~\cite{Bertolami2008,Bertolami2010,Francisco2012,Rievers2011,Turyshev2012}. The reason this effect was first detected on the Pioneers, but not on the Voyagers, is the use on the formers of spin-stabilisation and, as such, long periods without manoeuvrers, which allows for high precision in the trajectory determination. These conditions were reproduced in the Cassini Solar Conjunction experiment, which not only allowed for the best determination of the $\gamma$ parameter of the parametrised post-Newtonian (PPN) formalism so far~\cite{Bertotti2003}, but also led to the detection of a similar anomalous acceleration of thermal origin, just like the Pioneer spacecraft~\cite{Bertolami2014}.

The New Horizons mission has undergone some ``hibernation'' periods, where no thruster was fired. Therefore, it is very likely that a similar thermal origin acceleration might show up in the radiometric data. Indeed, this has been reported in Ref.~\cite{Rogers2014} for the period between February 2008 and May 2013. This detection leads to the necessity of comparing any such acceleration with a computation of the thermally induced acceleration.

We use the already proven pointlike source method, briefly described in section~\ref{sec:Source_method}, to the New Horizons spacecraft and its Pluto bound mission (\textit{cf.} section~\ref{sec:NH_Model}). Following the same procedure applied to the Pioneers and Cassini, we obtain a predicted range for the thermally induced acceleration and discuss the expected thermal effects on the attitude of the spacecraft, as discussed in sections~\ref{sec:NH_Model} and~\ref{sec:Results_Discussion}. Details of the attitude modelling are presented in the appendix.

\section{Pointlike Source Method}
\label{sec:Source_method}

\subsection{Motivation}
\label{subsec:Source_method_motivation}

The pointlike source method, first discussed in Ref.~\cite{Bertolami2008}, is an approach that maintains a high computational speed and a broad degree of flexibility, allowing for an easy analysis of different contributions and scenarios.

The method was designed to keep all the physical features of the problem easily scrutinisable. A battery of test cases can be performed to test the robustness of the result and ensure that the simplicity and transparency were not achieved at the expense of accuracy~\cite{Bertolami2010}.

One important feature of this method is its ability to consider parameters involving a large degree of uncertainty, which may arise from unavailability of accurate engineering data or lack of knowledge about changes due to the extended periods of degradation in space. By assigning a statistical distribution to each parameter, based on the available information, and generating a large number of random values, a Monte Carlo simulation can be used to generate a probability distribution for the final result~\cite{Francisco2012,Bertolami2014}.

The fact that this method was already used to deal with spacecraft thermal emissions in the context of the Pioneer anomaly and the Cassini solar conjunction experiment, producing results that are in agreement with radiometric navigation data and with the ones obtained through finite-element models~\cite{Rievers2011,Turyshev2012}, is a clear indication of its reliability and robustness.

\subsection{Radiative Momentum Transfer}
\label{subsec:RadMomTransf}

Before considering the particular problem at hand, it is useful to briefly review the physical formulation behind the pointlike source method.

The key feature of this method is the distribution of a small number of criteriously placed pointlike sources that model the thermal radiation emissions of the spacecraft. Typically, Lambertian sources are used to model surface emissions, although other types of sources to model particular objects are possible.

All the subsequent formulation of emission and reflection is made in terms of the Poynting vector field. For instance, the time-averaged Poynting vector field for a Lambertian source located at $\mathbf{x}_0$ is given by
\begin{equation}
    \label{eq:lambertian_source}
    \mathbf{S}_\mathrm{Lamb}(\mathbf{x})=\frac{W}{\pi ||\mathbf{x}-\mathbf{x}_0||^2} \, \mathbf{n} \cdot \frac{\mathbf{x}-\mathbf{x}_0 }{ ||\mathbf{x}-\mathbf{x}_0||}  \frac{\mathbf{x}-\mathbf{x}_0 }{ ||\mathbf{x}-\mathbf{x}_0||},
\end{equation}
where $W$ is the emissive power, and $\mathbf{n}$ is the surface normal.

The amount of power illuminating a given surface $W_\mathrm{ilum}$ can be obtained by computing the Poynting vector flux through the illuminated surface $S$, given by the integral
\begin{equation}
    \label{eq:energy_flux}
    E_\mathrm{ilum} = \int_S \mathbf{S} \cdot \mathbf{n}_\mathrm{ilum}~ dA,
\end{equation}
where $\mathbf{n}_\mathrm{ilum}$ is the normal vector of the illuminated surface.

The absorbed radiation transfers its momentum to the surface, yielding a \emph{radiation pressure} $p_\mathrm{rad}$ given by the power flux divided by the speed of light, for an opaque unit surface. There is also a radiation pressure on the emitting surface but with its sign reversed. If there is transmission (\textit{i.e.}, the surface is not opaque) the pressure is multiplied by the absorption coefficient. As for reflection, we see in the next two sections that it is treated as a re-emission of a part of the absorbed radiation.

Integrating the radiation pressure on a surface allows us to obtain the force and, dividing by the mass of the spacecraft ($m_\mathrm{SC}$), its acceleration
\begin{equation}
    \label{eq:force_integration}
    \mathbf{F}_\mathrm{th} = \int_S \frac{\mathbf{S} \cdot \mathbf{n}_\mathrm{ilum} }{ c} \frac{\mathbf{S} }{ ||\mathbf{S}||} dA \Rightarrow \mathbf{a}_\mathrm{th} = \frac{1 }{ m_\mathrm{SC}} \mathbf{F}_\mathrm{th}.
\end{equation}

The specific procedure to perform this integration deserves a discussion on its own. To determine the force produced by radiation of the emitting surface, the integral should be taken over a closed surface encompassing the latter. Equivalently, the force exerted on an illuminated surface requires an integration surface that encompasses it. Furthermore, considering a set of emitting and illuminated surfaces implies a proper accounting of the effect of shadows cast by each other, which must be subtracted from the estimated force on the emitting surface. It is then possible to read the thermally induced acceleration directly.

\subsection{Reflection Modelling -- Phong Shading}
\label{subsec:Phong}

The inclusion of reflections in the model is achieved through a method known as \emph{Phong Shading}, a set of techniques and algorithms commonly used to render the illumination of surfaces in three-dimensional computer graphics~\cite{Phong1975}.

This method is composed of a reflection model, which includes diffusive and specular reflection, known as the \emph{Phong reflection model}, and an interpolation method for curved surfaces modelled as polygons, known as \emph{Phong interpolation}.

The Phong reflection model is based on an empirical expression that gives the illumination value of a given point in a surface, $I_\mathrm{p}$, as
\begin{equation}
    \label{eq:phong_refl_mod}
    I_\mathrm{p} = k_\mathrm{a} i_\mathrm{a} + \sum_{m \in \text{lights}} \left[k_\mathrm{d} (\mathbf{l}_m \cdot \mathbf{n})i_\mathrm{d} + k_\mathrm{s} (\mathbf{r}_m \cdot \mathbf{v})^{\alpha} i_\mathrm{s} \right],
\end{equation}
where $k_\mathrm{a}$, $k_\mathrm{d}$, and $k_\mathrm{s}$ are the ambient, diffusive, and specular reflection constants, and $i_\mathrm{a}$, $i_\mathrm{d}$, and $i_\mathrm{s}$ the respective light source intensities. The vector quantities are the direction of the light source $m$, $\mathbf{l}_m$, the surface normal, $\mathbf{n}$, the direction of the reflected ray, $\mathbf{r}_m$, and the direction of the observer, $\mathbf{v}$. Finally, $\alpha$ is a ``shininess'' constant, and the higher its value, the more mirror-like is the surface.

To use this formulation to resolve a physics problem, some constraints should be taken into account. The ambient light parameters $k_\mathrm{a}$ and $i_\mathrm{a}$, while useful in computer graphics, are not relevant in our problem, since they parametrise the reflection behaviour relative to a background radiation source. Also, the intensities $i_\mathrm{d}$ and $i_\mathrm{s}$ should be the same, since the diffusive and specular reflection are relative to the same radiation sources.

This method provides a simple and straightforward way to model the various components of reflection, as well as a more accurate accounting of the thermal radiation exchanges between the surfaces on the spacecraft. In principle, there is no difference between the treatment of infrared radiation, in which we are interested, and visible light, for which the method was originally devised, allowing for a natural wavelength dependence of the above material constants.

Given the presentation of the thermal radiation put forward in subsection~\ref{subsec:RadMomTransf}, the Phong shading methodology was adapted from a formulation based on \emph{intensities} (energy per unit surface of the projected emitting surface) to one based on the energy flux per unit surface (the Poynting vector).

\subsection{Computation of Reflection}
\label{subsec:reflection_computation}

Using the formulation outlined in subsection~\ref{subsec:Phong}, the diffusive and specular components of reflection can be separately computed in terms of the Poynting vector field. The diffusive component of the reflection radiation is given by
\begin{equation}
    \label{eq:diffusive_reflection}
    \mathbf{S}_\mathrm{rd}(\mathbf{x},\mathbf{x}')=\frac{k_\mathrm{d} |\mathbf{S}(\mathbf{x}')\cdot \mathbf{n}| }{ \pi ||\mathbf{x}-\mathbf{x}'||^2} \, \mathbf{n} \cdot (\mathbf{x}-\mathbf{x}') \, \frac{\mathbf{x}-\mathbf{x}' }{ ||\mathbf{x}-\mathbf{x}'||},
\end{equation}
where $\mathbf{x}'$ is a point on the reflecting surface. Conversely, the specular component reads

\begin{equation}
    \label{eq:specular_reflection}
    \mathbf{S}_\mathrm{rs}(\mathbf{x},\mathbf{x}')=\frac{k_\mathrm{s} |\mathbf{S}(\mathbf{x}')\cdot \mathbf{n}| }{ \frac{2 \pi }{ 1+ \alpha} ||\mathbf{x}-\mathbf{x}'||^2} \, [\mathbf{r} \cdot (\mathbf{x}-\mathbf{x}')]^{\alpha} \, \frac{\mathbf{x}-\mathbf{x}' }{ ||\mathbf{x}-\mathbf{x}'||}.
\end{equation}
In both cases, the reflected radiation field depends on the incident one, $\mathbf{S}(\mathbf{x}')$, and on the diffusive and specular reflection coefficients $k_\mathrm{d}$ and $k_\mathrm{s}$, respectively. Adding up Eqs.~\eqref{eq:diffusive_reflection} and~\eqref{eq:specular_reflection} completely defines the reflected radiation field. From the emitted and reflected vector fields, the irradiation of each surface is computed and, from that, a calculation of the acceleration can be performed through Eq.~\eqref{eq:force_integration}.

When modelling the actual vehicle, and once the radiation source distribution is decided, the first step is to compute the emitted radiation field and the respective force exerted on the emitting surfaces. This is followed by determining which surfaces are illuminated, and computations of the force exerted on them by the radiation. At this stage, we obtain a figure for the thermal force without reflections. The reflection radiation field is then computed for each surface, and subject to the same steps as the initially emitted field, leading to a determination of thermal force with one reflection.

This method can, in principle, be iteratively extended to as many reflection steps as desired, considering the numerical integration algorithms and available computational power. If deemed necessary, each step can be simplified through a discretisation of the reflecting surface into pointlike reflectors.

\section{New Horizons Model}
\label{sec:NH_Model}

\subsection{Physical Setup}

Before modelling the emissions from the spacecraft's surfaces, these must be described by a geometric model. It has been shown before that this model only needs to describe the basic features, whereas small geometric details, when compared to the overall dimensions, do not significantly affect the results~\cite{Bertolami2008,Bertolami2010}.

\begin{figure}[!htb]
  \centering
  \includegraphics[width=0.8\columnwidth,clip=true,trim=2cm 6cm 2cm 3cm]{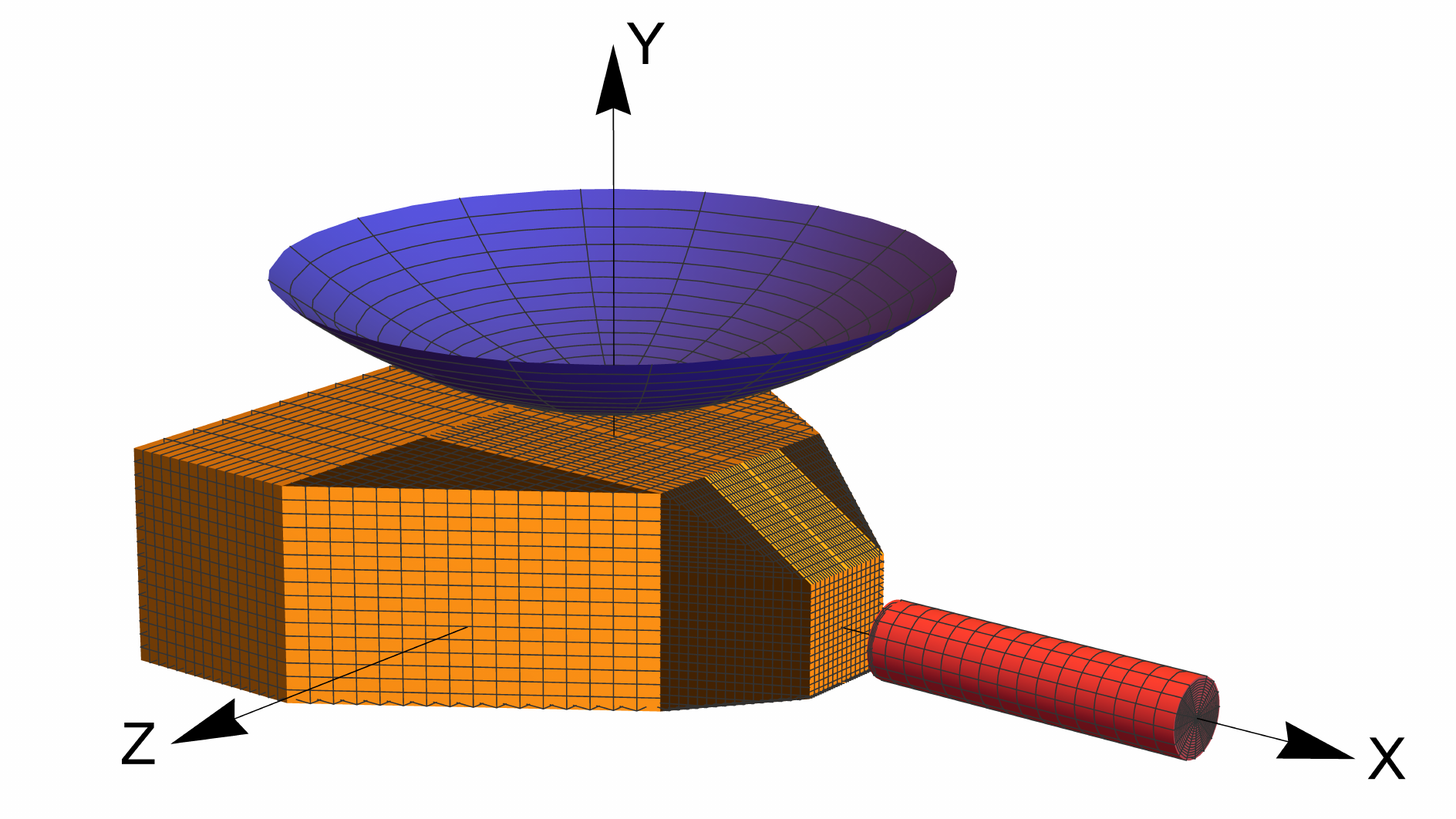}
  \caption{Model of the New Horizons used for the analysis, showing the antenna, main body, and RTG.}
  \label{fig:NewHorizonsSpacecraft_Model}
\end{figure}

\begin{figure}[!htb]
  \centering
  \includegraphics[width=0.53\columnwidth]{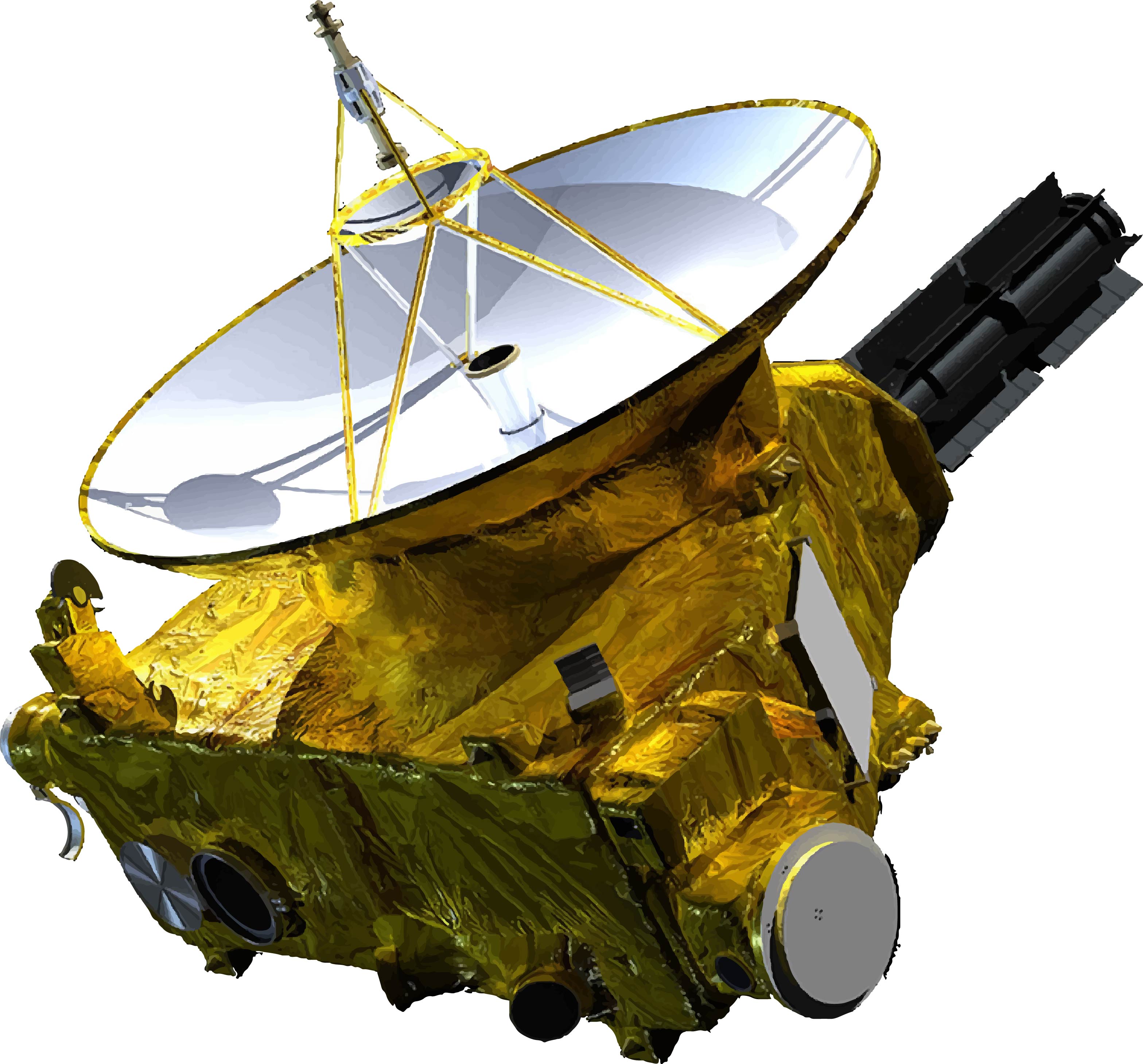}
  \caption{Artistic view of the New Horizons (adapted from~\cite{NASA2015}).}
  \label{fig:NewHorizons_Artist}
\end{figure}

The New Horizons spacecraft could be described, for the purposes of our computations, as a polyhedron shaped main body with a single RTG attached laterally and a main communication antenna on top of the main body. A schematics of this model is shown in Fig.~\ref{fig:NewHorizonsSpacecraft_Model}, created from the available specifications~\cite{NewHorizons_Model}. An artistic depiction of the spacecraft is shown in Fig.~\ref{fig:NewHorizons_Artist} for comparison.

The spacecraft's main body houses most of the electronics and payload instruments. With a length of about $2\,\unit{m}$ ($X$ direction on Fig.~\ref{fig:NewHorizonsSpacecraft_Model}), the height ($Y$ axis of the figure) varies from about $0.7\,\unit{m}$ to around $0.3\,\unit{m}$. It has a width of about $2\,\unit{m}$ in its widest point, and just $0.3\,\unit{m}$ in the narrowest one. As can be seen on the figure, the body is symmetric with respect to the $X$-$Y$ plane. Moreover, it is almost fully covered by multilayer thermal blankets (MLI) to maintain stable its temperature. These thermal blankets also influence the reflection of the radiation.

Thermal blankets also coat the back of the main antenna dish, which is used to provide $600\,\unit{bps}$ data downlink at $36\,\unit{AU}$. This high gain antenna is about $2.1\,\unit{m}$ in diameter, and has a depth of $0.4\,\unit{m}$. A support structure links it to the top of the main body, with a separation between the two of $0.12\,\unit{m}$.

At launch, the New Horizon spacecraft weighed $478\,\unit{kg}$, of which $77\,\unit{kg}$ was the hydrazine propellant and helium pressurant~\cite{Guo2008}. Of that propellant, it was estimated that $47\,\unit{kg}$ would still be available after the Pluto encounter.

Depending on the phase of the mission, New Horizons uses different attitude control modes. During transit the spacecraft is spin-stabilised, with its main antenna pointing to Earth, and only changing to three-axis control mode during some science observations and manoeuvrers~\cite{Fountain2008}. A trajectory determination accurate enough to detect the kind of subtle non-gravitational accelerations arising from thermal effects should only be available for the periods in which the spacecraft is spin-stabilised, since these measurements usually require the absence of manoeuvres.

\subsection{Power Supply}
\label{subsec:Power_Supply}

As previously stated, space missions that go beyond Mars rely on RTGs for power supply. New Horizons is no exception, having a single F-8 RTG, loaded with $9.75\,\unit{kg}$ of plutonium-238 dioxide~\cite{Ottman2006}. The RTG, measuring about $1\,\unit{m}$ and with a diameter of $0.3\,\unit{m}$ (not accounting for the cooling fins), is linked to the main body through an adaptor collar and a support structure, which provides both thermal and electric isolation~\cite{Fountain2008}.

Plutonium has a half-life of 87.7 years, and the thermal output generated was $3948\,\unit{W}$ at launch. The time evolution can be expressed, with $t$ in years, as
\begin{equation}
    \label{eq:thermal_power_evolution}
    W_\mathrm{thermal}^\mathrm{RTG}(t) = 3948\times e^{-t\,\log(2)/87.7}\,\unit{W},
\end{equation}
where $t = 0$ corresponds to the time of launch.

Of the thermal output produced, only a part is converted into electrical power, according to the efficiency of the thermocouples. Furthermore, this efficiency decreases over time faster than the Plutonium thermal output. For instance, at launch the RTG fed the system with $246\,\unit{W}$, whereas at Pluto only a little over $200\,\unit{W}$ were available. To get an evolution of the electrical power produced, a fit to the data of Ref.~\cite{Fountain2008} was made. Assuming a type of evolution similar to the thermal output, but with a smaller half-life, we have
\begin{equation}
    \label{eq:electric_power_evolution}
    W_\mathrm{electric}^\mathrm{RTG}(t) = 241.3\times e^{-t\,\log(2)/39.1}\,\unit{W}.
\end{equation}

We can assume that the electric power, Eq.~\eqref{eq:electric_power_evolution}, is to be introduced in Eq.~\eqref{eq:lambertian_source}, as the power emitted by the main bus ($W_\mathrm{Bus}(t) = W_\mathrm{electric}^\mathrm{RTG}(t)$), since sooner or later most of that power is transformed into heat. The power emitted by the RTG, in order to respect the energy equilibrium, is the balance of both the thermal and electric (\textit{i.e.} $W_\mathrm{RTG}(t) = W_\mathrm{thermal}^\mathrm{RTG}(t) - W_\mathrm{electric}^\mathrm{RTG}(t)$).

\subsection{Thermal Emissions Modelling}
\label{sec:Initial_Study}

The next step in the construction of our model is to determine the number and distribution of sources that should be considered. Since the RTG is the component with higher emitting power, we first turn our attention to this component.


Thermal radiation from the RTG is emitted in all directions, since it is a cylindrical body. However, we are only interested in the radiation reflected by the main antenna dish. The remaining radiation force is either symmetric, cancelling it out, or perpendicular to the spacecraft spin axis. Using Eqs.~\eqref{eq:lambertian_source}~and~\eqref{eq:energy_flux}, we calculate how the energy flux varies with the emitted power.

To model the cylindrical shape of the RTG with flat surface Lambertian sources we use an increasing number of sources in the azimuthal direction of the cylinder; starting with a pair of sources, we move to 4, 8, 16 and finally 32 sources in the azimuthal direction, for each axial section. Along the length of the RTG, we consider four and eight axial sections.

\begin{figure}[!htb]
  \centering
  \includegraphics[width=1\columnwidth]{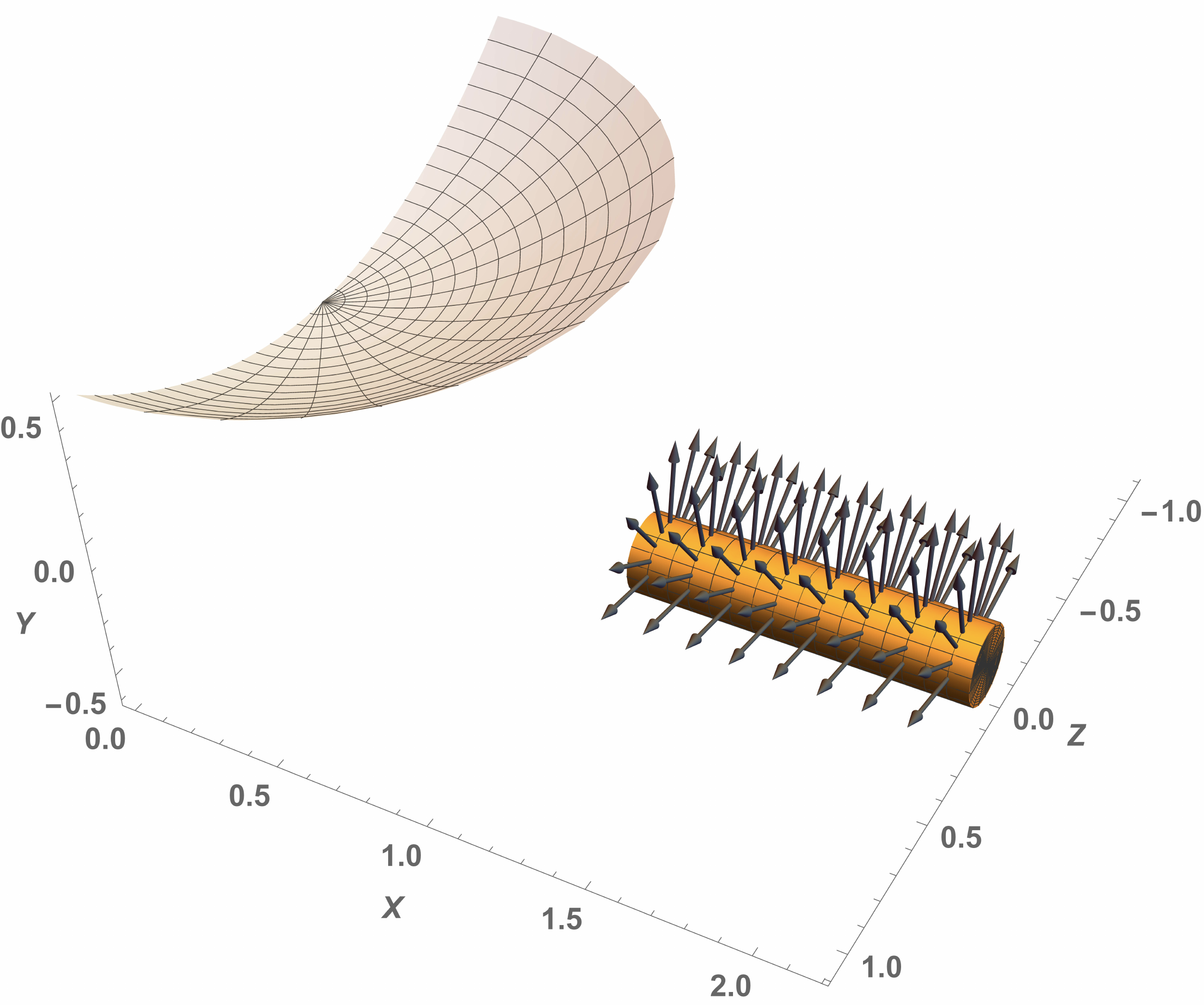}
  \caption{Distribution of the eight times sixteen Lambertian sources on the RTG (only sources in the upper half are shown). The object on top is a part of the high gain antenna, illuminated by the displayed radiation sources.}
  \label{fig:RTG_Sources}
\end{figure}

Running a convergence analysis for the energy flux, we find that the difference from four to eight axial sections influences the energy flux by only about $0.02\%$, which is clearly negligible for the expected level of accuracy. However, in the azimuthal direction, the difference between the 2 and 16 source model is, as could be expected, larger at $\approx 2\%$. Still, and since the change from a 16-sided polygon to a 32-sided one only makes the results change by 0.018\%, we decided that the model with 16 sources in the azimuthal direction by four axial sections gave the best balance between computational effort and accuracy. A representation of the selected configuration can be seen in Fig.~\ref{fig:RTG_Sources}.


Apart from the RTG, the main body is the only other source of radiation. However, each surface emits in a particular direction, with a different intensity, which has to be taken into account. Therefore, to have a correct distribution of the emitted power, the total power ($W_\mathrm{Bus}(t)$) is divided by each surface according to its area fraction ($f_{A_i} = A_i / A_\mathrm{total}$, where $A_i$ represents the surface's area and $A_\mathrm{total}$ is the total body area). This is a reasonable assumption since the MLI covers all the body and evens the temperature between surfaces.

The main body either illuminates the main antenna, or irradiates to deep space. The former gives rise to a radiation momentum as discussed in section~\ref{sec:Source_method}. The radiation to deep space has a simpler treatment.

\begin{figure}[!htb]
  \centering
  \includegraphics[width=1\columnwidth]{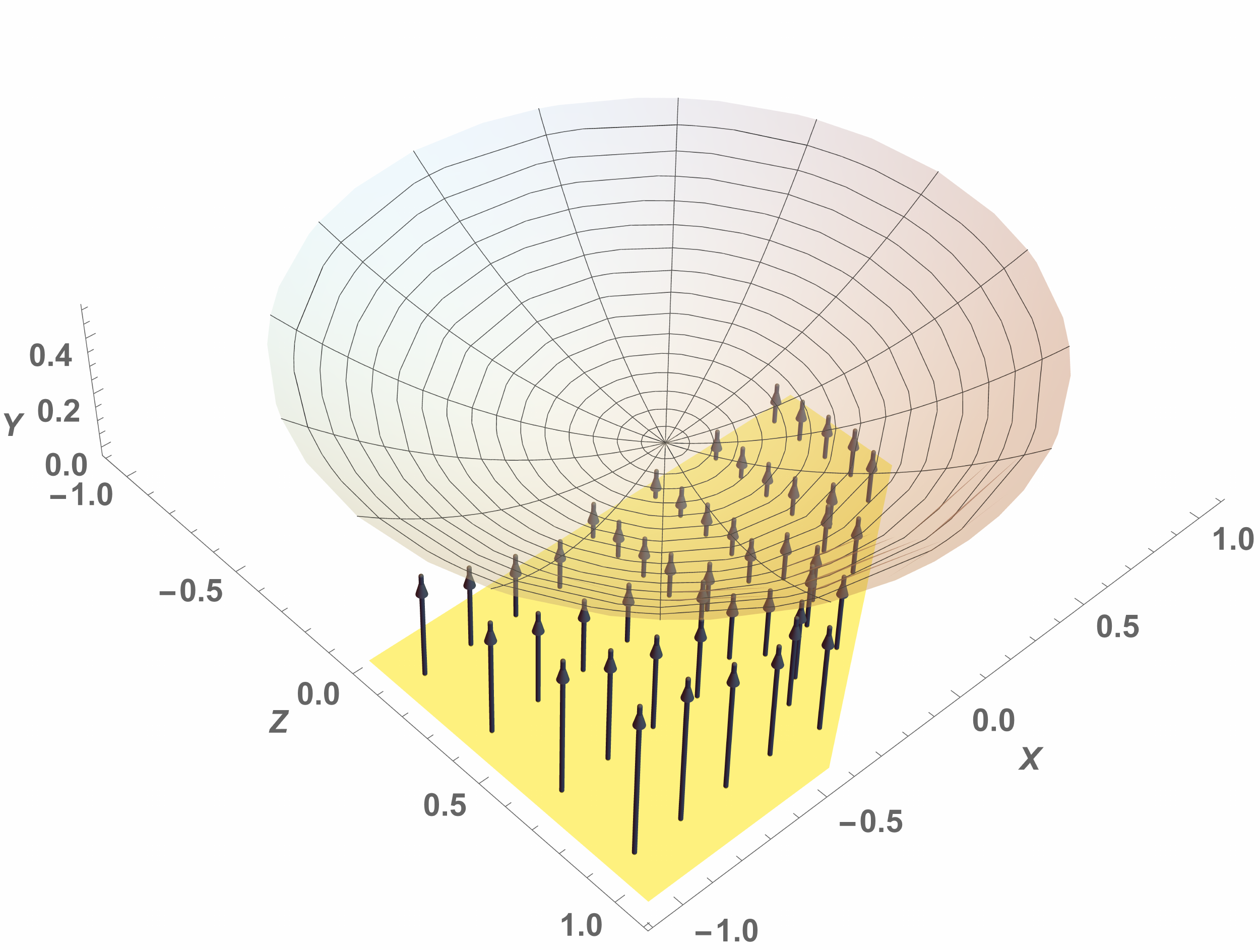}
  \caption{Distribution of the sixteen Lambertian sources (represented by the arrows) on the top wall of the main body (only half of the wall is shown due to symmetry).}
  \label{fig:TopWall_Sources}
\end{figure}

The surface with the biggest area, which shines on the antenna, is the top wall. Modelling it with a different number of sources, we get a difference of 0.04\% between using a single source or 16 sources. The difference is just 0.001\%, between eight and 16 sources. We thus decided not to pursue the division of the top wall and use 16 sources, which is depicted in Fig.~\ref{fig:TopWall_Sources}.

A similar process was followed for the remaining surfaces that irradiate the main antenna. For those, only eight sources per surface were used. We found it sufficient since differences between four and eight sources were less than 0.0007\%, and all combined represented an energy flux of less than 3\% of the overall power.


The bottom part of the main body does not illuminate any surface and radiates to deep space. This is also true for some of the side walls. However, because the spacecraft is spin-stabilised, only the bottom part has any real contribution for a thermal acceleration. Considering each surface as a Lambertian emitter, the radiation force can simply be given by
\begin{equation}
    \label{eq:radiation_force}
    \mathbf{F}_{\mathrm{rad}_i} = \frac{2}{3}\frac{f_{A_i}\,W_\mathrm{Bus}}{c}\,\mathbf{n}_i.
\end{equation}

\section{Results and Discussion}
\label{sec:Results_Discussion}

\subsection{Baseline Scenarios}
\label{subsec:Scenarios}

Before performing a comprehensive statistical analysis, it is important to have a few figures of merit and to gain some intuition on how the different parameters affect the final result.

There are two instants in the New Horizons mission for which we have a good idea about most of the necessary parameters: the launch and the encounter with Pluto. Both the thermal output of the RTG and the electrical power available for the bus were maximum at launch. We know that immediately after launch, any thermally generated acceleration is masked by the solar pressure effect, since it only drops low enough to allow the detection of thermal accelerations beyond the asteroid belt. This fact is known from the original detection of the Pioneer anomaly \cite{Anderson2002}. Still, it might be a useful exercise to consider the power output from New Horizons right after launch so as to give us an upper bound for any thermal acceleration.

The first hypothesis is to assume no reflection at all. This would mean that the acceleration is only due to the emitted radiation. Using the values and the formulas for the mass and power discussed above ($m_\mathrm{NH} = 478\,\unit{kg}$, $W_\mathrm{Bus}(t = 0) = 240.4\,\unit{W}$, $W_\mathrm{RTG}(t = 0) = 3707\,\unit{W}$), the resulting acceleration is
\begin{equation}
    \label{eq:acc_launch_no_reflection}
    \mathbf{a}_\mathrm{{Hyp\,1}} = \left(
    - 5.4\, \mathbf{e}_X
    + 8.5\, \mathbf{e}_Y
    +   0\, \mathbf{e}_Z
    \right)
    \times 10^{-10}~\unit{m/s^2}.
\end{equation}
From this simple calculation we confirm what was already expected; the acceleration is null on the $Z$ direction, because of the spacecraft symmetry in the $XY$ plane. An interesting fact is the resulting acceleration in the $X$ direction. Although it is not negligible, when compared to the $Y$ direction, it would only be relevant if the spacecraft was not spinning.

The next step is to add reflection to the previous result. The mass and power are already defined, whereas the reflection coefficients (diffusive and specular) and the shininess constant have to be set. Since the spacecraft is fully covered with the MLI, the total reflection can be considered high (up to 95\% according to the New Horizons MLI manufacturer~\cite{DUNMORE_web2016}), but very diffusive and with a small shininess. Therefore, assuming $k_\mathrm{d} = 0.85$, $k_\mathrm{s} = 0.1$, and $\alpha = 3$, our model yields
\begin{equation}
    \label{eq:acc_launch_hyp2}
    \mathbf{a}_\mathrm{{Hyp\,2}} = \left(
    -  7.2\, \mathbf{e}_X
    + 14.9\, \mathbf{e}_Y
    +    0\, \mathbf{e}_Z
    \right)
    \times 10^{-10}~\unit{m/s^2}.
\end{equation}
Here the same holds, the null component for $\mathbf{e}_Z$ and the non-null for $\mathbf{e}_X$ and $\mathbf{e}_Y$. The relevant fact, however, is the considerable increase of the absolute values brought by the reflection.

Although we could move to the parametric study from here, both the value of the total reflection as well as the distribution between diffusive and specular components are not very well defined and thus required a simple analysis. Considering a new hypothesis of total reflectance of 60\%, which is a typical value for aluminised kapton (the type of MLI used in New Horizons) with the aluminium on the inside~\cite{Martinez2016}, and a 75\% diffusive component, the new coefficients would be $k_\mathrm{d} = 0.45$ and $k_\mathrm{s} = 0.15$. With these values, the acceleration is now 
\begin{equation}
    \label{eq:acc_launch_hyp3}
    \mathbf{a}_\mathrm{{Hyp\,3}} = \left(
    -  6.4\, \mathbf{e}_X
    + 12.5\, \mathbf{e}_Y
    +    0\, \mathbf{e}_Z
    \right)
    \times 10^{-10}~\unit{m/s^2}.
\end{equation}
Another possibility is to have about the same total reflection as before (65\%), but with a higher diffusive component (85\%). This last case yields
\begin{equation}
    \label{eq:acc_launch_hyp4}
    \mathbf{a}_\mathrm{{Hyp\,4}} = \left(
    -  6.6\, \mathbf{e}_X
    + 12.9\, \mathbf{e}_Y
    +    0\, \mathbf{e}_Z
    \right)
    \times 10^{-10}~\unit{m/s^2},
\end{equation}
for coefficients $k_\mathrm{d} = 0.55$ and $k_\mathrm{s} = 0.10$. This hypothesis is the starting point for the parametric study.

\subsection{Parametric Study}
\label{subsec:parametric_study}

A parametric study, using the Monte Carlo method, allows for an analysis on a large range of values, statistically distributed, for all parameters involved.

The mass is a well known parameter, at launch and at the encounter with Pluto, and thus any statistical variation can be discarded at those two instances. However, as we aim to perform an exponential fit of the data, we need a third point. Thus, we decide to compute the acceleration at half of the trip. For that moment, we can assume that at least half of the spent propellant to reach Pluto was used by then. This is in agreement with the data from Ref.~\cite{Rogers2014}, and thus the value of $461\,\unit{kg}$ is used.

The spacecraft energy, although defined by Eqs.~\eqref{eq:thermal_power_evolution} and~\eqref{eq:electric_power_evolution}, still has some uncertainty associated with it. Therefore, the decision was to use a normal distribution for each value. However, as there are values with higher and smaller uncertainties, we defined different standard deviations. For low confidence values, $\sigma$ is 3.5\% of the obtained value with the power equation, while for the others $\sigma$ is 1.5\%.

The reflection coefficients are the least known parameters, and as such, we decided to use an uniform distribution. Therefore, the diffusive component, $k_\mathrm{d}$, varies between between 0.55 and 0.95 of the total reflection, whereas the specular component is just $k_\mathrm{s} = k_\mathrm{total} - k_\mathrm{d}$. Furthermore, we included a decrease on the total reflection with time, to account for optical property degradation due to exposure to space, and defined a total reflectance of 0.8 at launch, 0.65 at half of the trip, and 0.5 when encountering Pluto.

\begin{figure}[!htb]
  \centering
  \includegraphics[width=1\columnwidth]{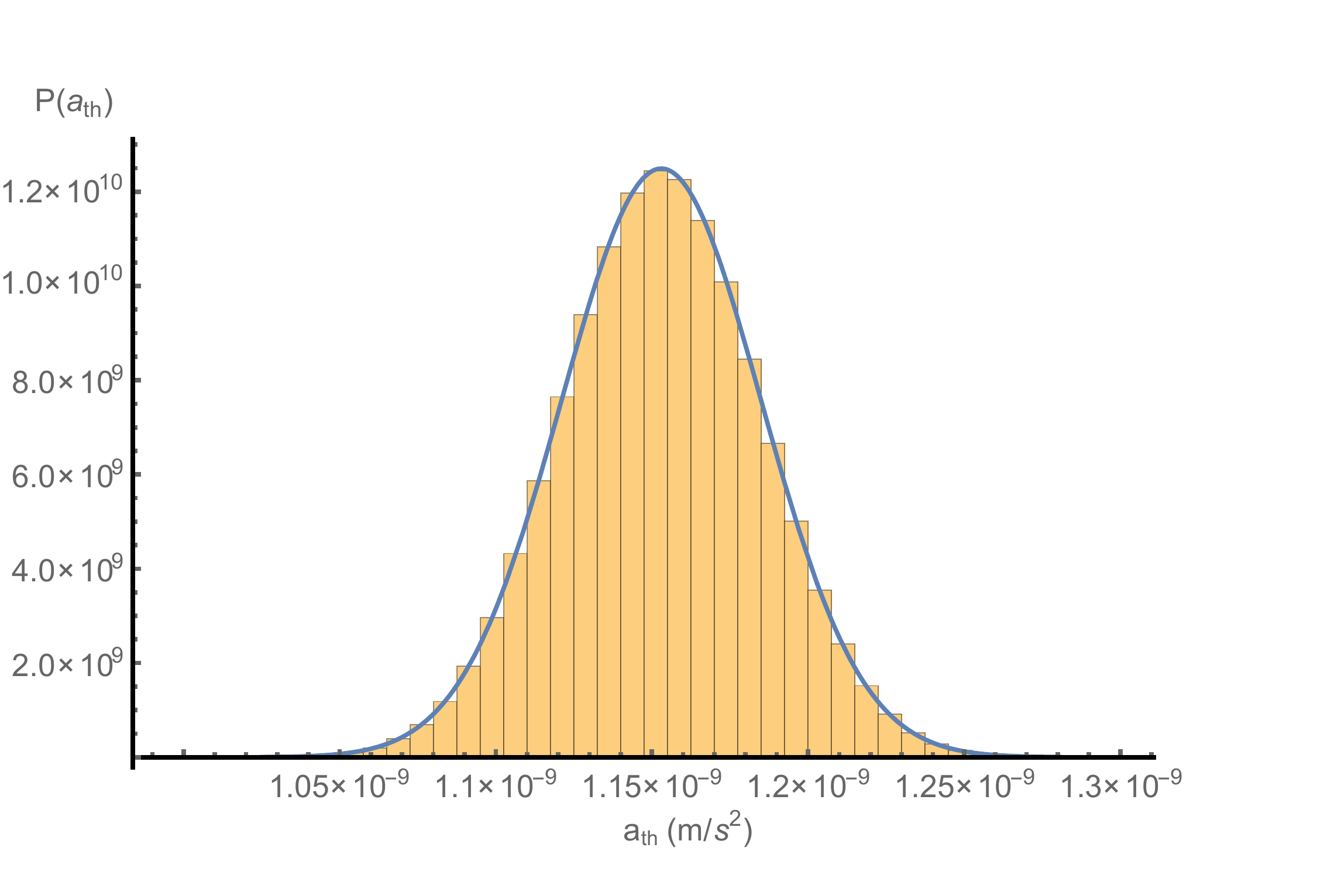}
  \caption{Resulting probability distribution for the thermal acceleration in the $Y$ direction at the time of encounter with Pluto, overlaid with the derived normal distribution.}
  \label{fig:ProbDensity_Encounter}
\end{figure}

Ater simulating $10^6$ hypothesis, the resulting probability density distribution for the time of encounter is shown in Fig.~\ref{fig:ProbDensity_Encounter}.
As can be seen, the distribution has a normal characteristic, and we can state that the $Y$ component of the thermal acceleration is
\begin{equation}
	\left(a_\mathrm{NewHorizons}\right)_Y = (11.5 \pm 0.3) \times 10^{-10}~\unit{m/s^2},
\end{equation}
with a $2\sigma$ uncertainty interval, for $t = 9\,\unit{years}$.

\begin{figure}[!htb]
  \centering
  \includegraphics[width=1\columnwidth]{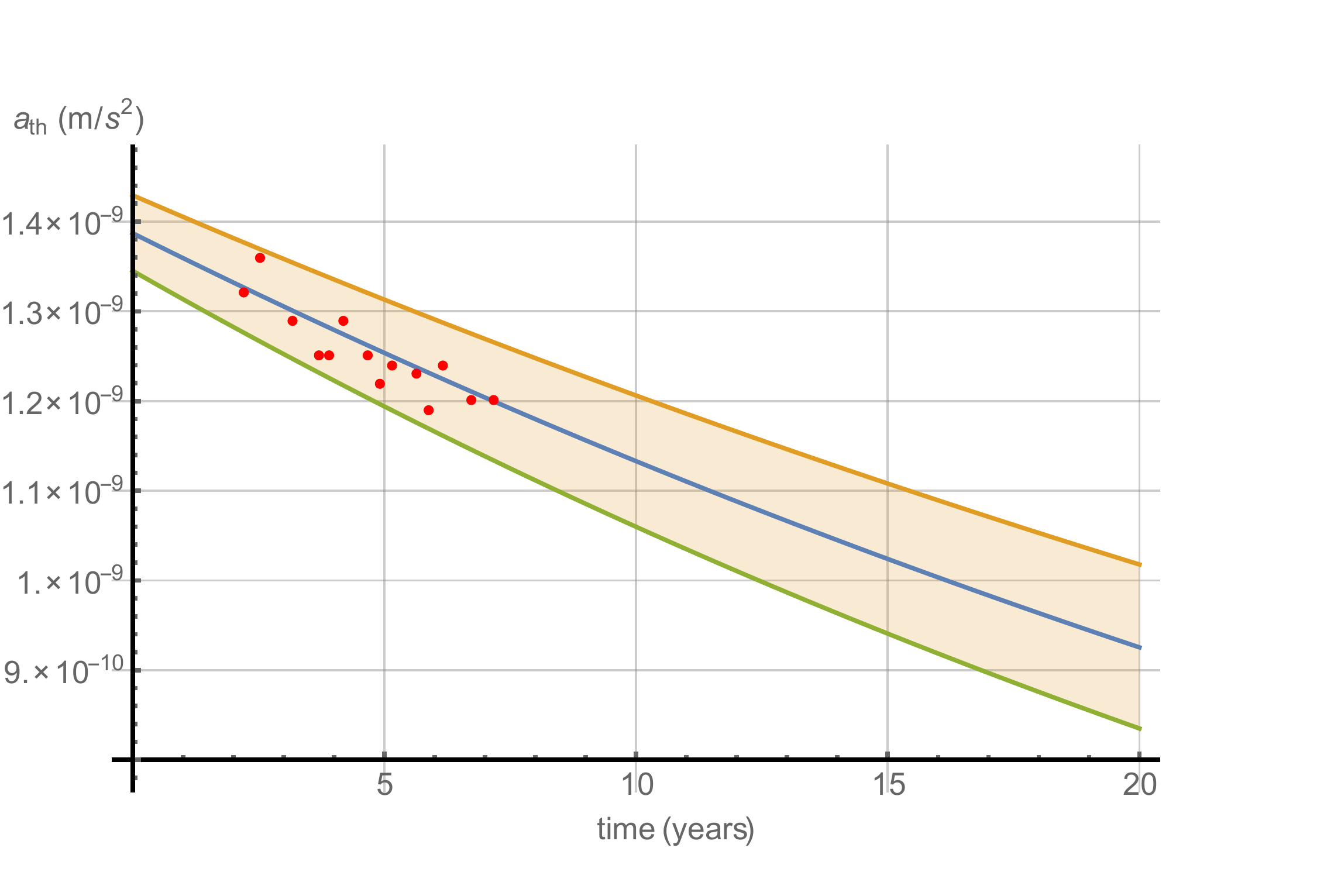}
  \caption{Exponential fit of the thermal acceleration along the $Y$ axis, and the $2\sigma$ confidence region around it, for a 20-year period after launch. As a note, Pluto was encountered after a nine-year trip. The dots represent data from Ref.~\cite{Rogers2014}.}
  \label{fig:ExpFit_Acceleration}
\end{figure}

After computing the value of the acceleration for the other two instants of time, the launch and middle of the mission, we can get an estimate for the evolution of the acceleration, which is shown in Fig.~\ref{fig:ExpFit_Acceleration} for a period of 20 years after launch.
It is clear that a thermal acceleration in the $Y$ direction is not negligible, and decreases as the RTG releases less energy. Furthermore, the acceleration estimates from Ref.~\cite{Rogers2014}, fall within our results, validating the parameters assumed for that period.

Although the value of the thermal acceleration is relatively small, it can have significant impact on the spacecraft trajectory, since the value accumulates over time. As an example, if we assume that after the Jupiter flyby, in February 2007, the spacecraft did not perform any trajectory corrections up to the encounter with Pluto (\textit{i.e.} it spent 6 years in hibernation), there would be a difference in its position of $\Delta Y \approx 2 \times 10^4\,\unit{km}$.

\subsection{Results of the Thermal Effects on Attitude}
\label{sec:attitude_results}

Since we know that the thermal force and its location at each area element, are sufficiently small for the error level of the problem, we can determine the thermal moment and estimate its effect on the attitude, by knowing the inertia tensor of the spacecraft and the location of its centre of mass. Since we know the thermal effects are small we can use a simple first order model (\textit{cf.} the appendix) to determine the attitude evolution in free fall, with or without spin stabilization.

The nominal spin rate of New Horizons is $\omega_0=5\,\unit{rpm}$~\cite{Fountain2008}, and the location of the centre of mass is the origin of the reference frame (in Fig.~\ref{fig:NewHorizonsSpacecraft_Model}). In this reference frame the principal moments of inertia are $A=I_{XX}=161.38\,\unit{kg/m^2}, B=I_{YY}=402.12\,\unit{kg/m^2}, C=I_{ZZ}=316\,\unit{kg/m^2}$~\cite{Fountain2017}.

Using Eq.~\eqref{eq:force_integration} and summing up the contributions from each element (Eq.~\eqref{eq:moment_general}), the resultant thermal moment at launch is
\begin{equation}
    \label{eq:moment_result}
    \mathbf{M}(t=0) \approx \mathbf{M}_Z = \left(
    + 5.76 \pm 0.25
    \right)
    \times 10^{-7}~\unit{N.m}.
\end{equation}
The other moment components are much smaller than $M_Z$; both $M_X$ and $M_Y$ are of the order of $10^{-12}~\unit{N.m}$, and can be safely discarded, as expected (\textit{cf.} Eq.~\eqref{eq:moment_particular}). Furthermore, these values are the upper bound of the moment.

\subsubsection{Spin Stabilised Case}

Substituting Eq.~\eqref{eq:moment_result} into Eq.~\eqref{eq:psi_nu}, we obtain the Euler angles as function of time for the New Horizons in passive spin mode. The maximum expected deviations $\psi,\nu$ and their rates $\dot{\psi},\dot{\nu}$ are approximately
\begin{align}
    |\psi|          &\lesssim \left(12.7 \pm 0.6 \right)\times 10^{-9}\,\unit{rad},\label{eq:estimate_psi}
    \\
    |\dot{\psi}|    &\lesssim \left(9.14 \pm 0.4 \right)\times 10^{-9}\,\unit{rad/s},\label{eq:estimate_dpsi}
    \\
    |\nu |          &\lesssim \left(17.9 \pm 0.8 \right)\times 10^{-9}\,\unit{rad},\label{eq:estimate_nu}
    \\
    |\dot{\nu}|     &\lesssim \left(9.14 \pm 0.4 \right)\times 10^{-9}\,\unit{rad/s}.\label{eq:estimate_dnu}
\end{align}
The effect produced is very small, with a frequency of about $10^{-8}\,\unit{rad/s}$, and is probably difficult to detect. The spacecraft's attitude control system is capable of providing spin axis attitude knowledge to better than $471\,\unit{\mu rad}$~\cite{Fountain2008}, about four orders of magnitude higher than the required level to detect the thermal effect. This wobble could possibly be detected from Earth during transmissions, since the signal intensity decays rapidly with the line-of-sight misalignment, an effect that might be detected in the communications from the spacecraft, although it seems to require a very high sensitivity.

The variation of the gain as a function of the direction in parabolic antennae can be approximately described (for $\theta$ smaller than $\theta_{3\unit{dB}}/2$) by $\Delta G(\theta) = - 24\theta \Delta\theta / \theta_{3\unit{dB}}^2$, where $\Delta G(\theta)$ is the gain variation at an angle $\theta$ from boresight in dB, $\theta_{3\unit{dB}}$ is the half power beamwidth, and $\Delta\theta$ is the angle deviation inducing the gain change~\cite{Maral2009}. In the direction of the boresight $\Delta G$ tends to zero, making it difficult to observe any effect. The maximum gain reduction is achieved when $\theta = \theta_{3\unit{dB}}/2$~\cite{Maral2009}. For this angle, using the beamwidth of the New Horizon's high gain antenna, $\theta_{3\unit{dB}} = 0.3^\circ$~\cite{DeBoy2004}, and for a deviation $\Delta\theta = \nu = 1.8\times 10^{-8}\,\unit{rad}$ (see Eq.~\eqref{eq:estimate_nu}) the gain reduction is $4.1\times 10^{-5}\,\unit{dB}$.

\subsubsection{Three-Axis Mode Case}

In the case of the three-axis control mode, using Eqs.~\eqref{eq:3_axis_mode} and~\eqref{eq:force_integration} the induced angular velocity around the $Z$ axis by the thermal effect is found to be, at launch,
\begin{equation}\label{eq:estimate-no-spin}
\delta\dot{\omega}_Z \simeq \left(18.2 \pm 0.8 \right)\times 10^{-10}\,\mathrm{rad/s}.
\end{equation}
Although the effect is also very small, it grows with time, so in principle, it could be detected. The spacecraft turns about $5\times10^{-6}\,\unit{rad}$ per day. However, in this mode the spacecraft does not spend a lot of time without manoeuvring unless it is set for that specific purpose, as, for instance, the Cassini spacecraft during the Solar conjunction experiment~\cite{Bertolami2014}. In this specific experiment, the spacecraft spent approximately 30 days without any manoeuvre. One could check for the largest possible accumulated rotation by assuming a cumulative effect throughout the whole period, leading to a deviation of $1.5 \times 10^{-4}\,\unit{rad}$. This would translate into a gain reduction of $0.34\,\unit{dB}$ in the direction of the maximum gain variation.

\section{Conclusions}
\label{sec:Conclusions}

In this work we have studied the forces due to thermal effects on the New Horizons spacecraft, and reached three main conclusions. First, we have set the already developed method for determining thermal accelerations using a radiation momentum, and the Phong shading model, to account for the thermal effects on the New Horizons spacecraft. Second, it is shown that the New Horizons has indeed a thermal induced acceleration in the Earth direction (to where the antenna is pointing) of the order of $(11.5 \pm 0.3) \times 10^{-10}~\unit{m/s^2}$, at the time of encounter with Pluto. Third, the thermal effect on the attitude is small, but it can possibly be detected in the communication signals, a suggestion that would constitute a direct measurement.

The acceleration estimate could be refined if more information on the reflective properties of the thermal blankets could be obtained. However, this would only make the confidence regions smaller and would not affect the overall conclusion

Now that we know there is a thermal acceleration present in New Horizons, and its estimated evolution, it would be interesting if NASA and the Johns Hopkins University Applied Physics Laboratory could look for this acceleration in the Doppler shifts on the tracking algorithms, for different and larger periods. They could also try to detect the small wobble that, in principle, can induce intensity variations during transmissions from the spacecraft with an estimated frequency. This would validate the method here described and further boost the confidence on the method proposed in Refs.~\cite{Bertolami2008,Bertolami2010,Francisco2012,Bertolami2014}.


\begin{acknowledgments}
    The work of A. Guerra is supported by the Funda{\c c}\~ao para a Ci\^encia e a Tecnologia (Portuguese Agency for Research) under fellowship Grant No. PD/BD/113536/2015. The work of F. Frederico was partially supported by FIRE-RS project, number SOE1/P4/E0437, funded by the Interreg Sudoe Programme and by the European Regional Development Fund (ERDF) at Faculdade de Engenharia da Universidade do Porto, Laboratório de Sistemas e Tecnologia Subaquática. The work of P.J.S. Gil was supported by FCT, through IDMEC, under LAETA, Grant No. UID/EMS/50022/2013.
\end{acknowledgments}

\appendix*

\section{Modelling the Thermal Effects on the Attitude of the Spacecraft}
\label{sec:Thermal_attitude}

\subsection{Thermal Moment Relative to the Centre of Mass}
\label{subsec:thermal_moment}

The New Horizons spacecraft has a principal moment of inertia axis along the direction of the antenna, axis $+Y$ in Fig.~\ref{fig:NewHorizonsSpacecraft_Model}, which is the nominal axis of rotation of the spacecraft~\cite{Fountain2008}. Furthermore, the spacecraft is reasonably symmetric about the $X$-$Y$ plane, as can be seen in section~\ref{sec:NH_Model}. Given this geometry, we assume the body reference frame of the figure has origin at the centre of mass, and is the principal reference frame with moments of inertia $A,B,C$.

Letting $\mathbf{F}$ be the resultant of the thermal radiation force, and $\mathbf{r}$ its position, with respect to the centre of mass, then the moment $\mathbf{M}$ can be written as
\begin{equation}
\label{eq:moment_general}
\mathbf{M}= \mathbf{r}\times \mathbf{F} = \sum_i \mathbf{r}_i \times \mathbf{F}_i,
\end{equation}
where the right hand side is the sum of the moments of each individual element at position $\mathbf{r}_i$ and with applied thermal force $\mathbf{F}_i$. Assuming that the force is approximately in the $X$-$Y$ plane we can expect that, due to the symmetry of the spacecraft, the moment is approximately
\begin{equation}
\label{eq:moment_particular}
\mathbf{M} = M \mathbf{e}_Z.
\end{equation}

\subsection{Rotation Dynamics Induced by Thermal Effects}
\label{subsec:thermal_rotation}

Depending on the mode of operation of the attitude control system, different conditions have to be considered. We focus on the spin stabilised and three-axis mode cases, since for the active spin mode it will not be possible to identify the effect of the thermal force.


\subsubsection{Spin Stabilised Case}
\label{subsec:spinsatibised_case}

In the case of the passive spin mode, the spacecraft is put to rotate around the principal $Y$ axis at nominal value $\omega_0$. The evolution of the rotational state is determined by the Euler equations~\cite{Wiesel1997}, which in the body reference frame take the form
\begin{subequations}
    \label{eq:Euler_eq}
    \begin{align}
    \label{eq:Euler_eqa}M_X &= A \dot{\omega}_X+(C-B)\omega_Y\omega_Z, \\
    \label{eq:Euler_eqb}M_Y &= B \dot{\omega}_Y+(A-C)\omega_X\omega_Z, \\
    \label{eq:Euler_eqc}M_Z &= C \dot{\omega}_Z+(B-A)\omega_X\omega_Y,
    \end{align}
\end{subequations}
The initial angular velocity of the spacecraft is $\bm{\omega}=\omega_0\mathbf{e}_Y$ with the $Y$ axis initially with fixed direction in space. Later the initial angular velocity is perturbed by the applied moment, and thus
\begin{equation}
\label{eq:spin_mode}
\bm{\omega} = \delta\omega_X \mathbf{e}_X + (\omega_0+\delta\omega_Y) \mathbf{e}_Y + \delta\omega_Z \mathbf{e}_Z,
\end{equation}
where the time dependent components $\delta\omega_i, i=X,Y,Z$ are initially zero, and are expected to be very small so that we can neglect higher order terms.

Recalling the thermal moment, Eq.~\eqref{eq:moment_general}, and noting that the New Horizons has a flattened body relative to the rotation axis $Y$, $B > A, C$, the perturbed components of the angular velocity are found to be
\begin{subequations}
    \label{eq:Euler_eq_sol}
    \begin{align}
    \label{eq:Euler_eq_sola}\delta\omega_X &= \alpha_1 \left[1-\cos(\Omega t) \right], \\
    \label{eq:Euler_eq_solb}\delta\omega_Y &= 0, \\
    \label{eq:Euler_eq_solc}\delta\omega_Z &= \alpha_2\sin(\Omega t),
    \end{align}
\end{subequations}
with
\begin{subequations}
    \label{eq:const}
    \begin{align}
    \label{eq:consta} \alpha_1 &= \frac{MR}{A\Omega}, \\
    \label{eq:constb} \alpha_2 &= \frac{M}{C\Omega}, \\
    \label{eq:constc} \Omega   &= \omega_0\sqrt{(B/A-1)(B/C-1)}, \\
    \label{eq:constd} R        &= \sqrt{\frac{A(B-C)}{C(B-A)}}.
    \end{align}
\end{subequations}
Since $B$ is the largest principal moment of inertia, $\Omega$ is real and has the same sign of $\omega_0$.

\begin{figure}[!ht]
    \centering
    \includegraphics[width=.55\columnwidth]{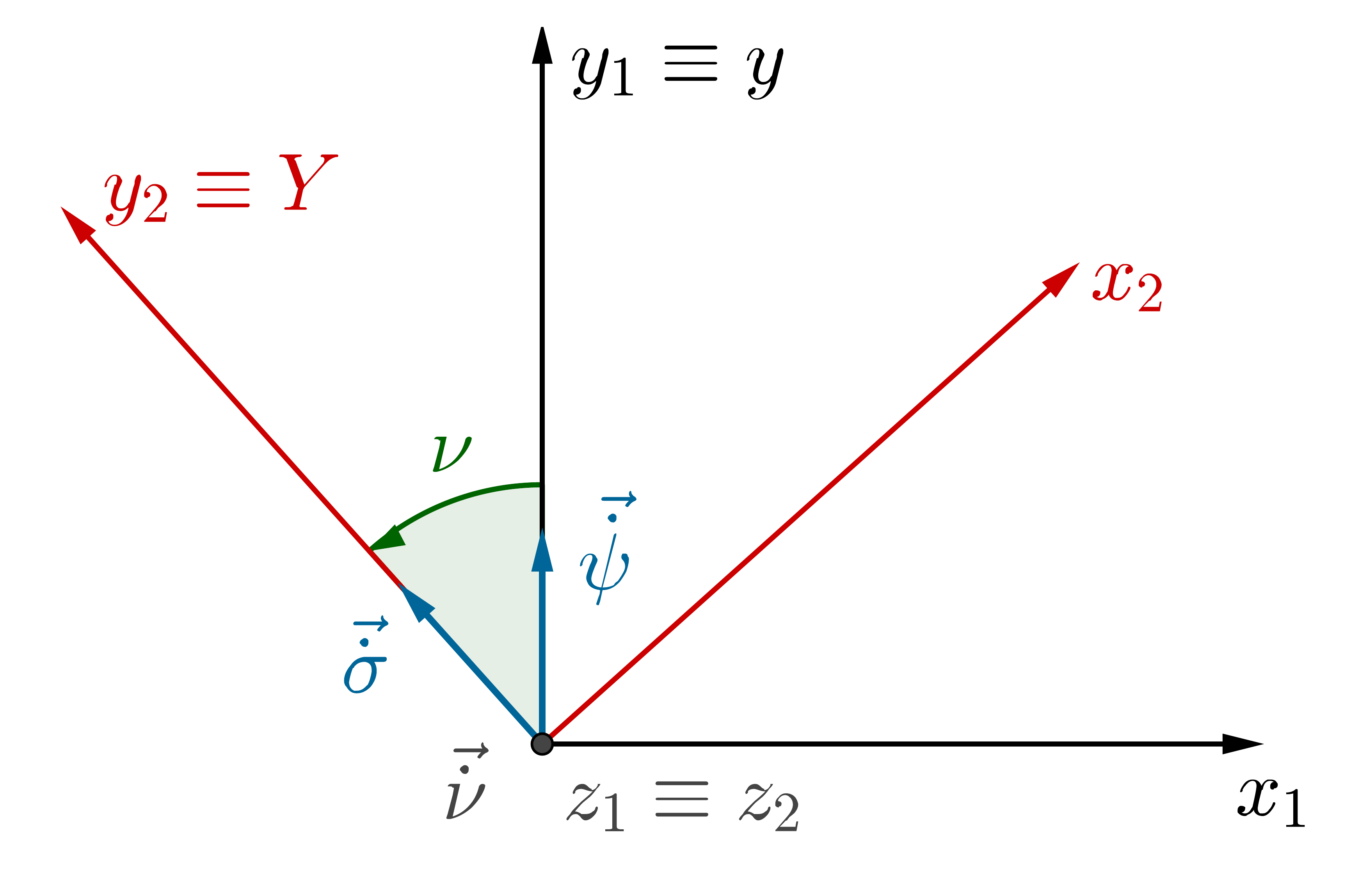}\hfill
    \includegraphics[width=.45\columnwidth]{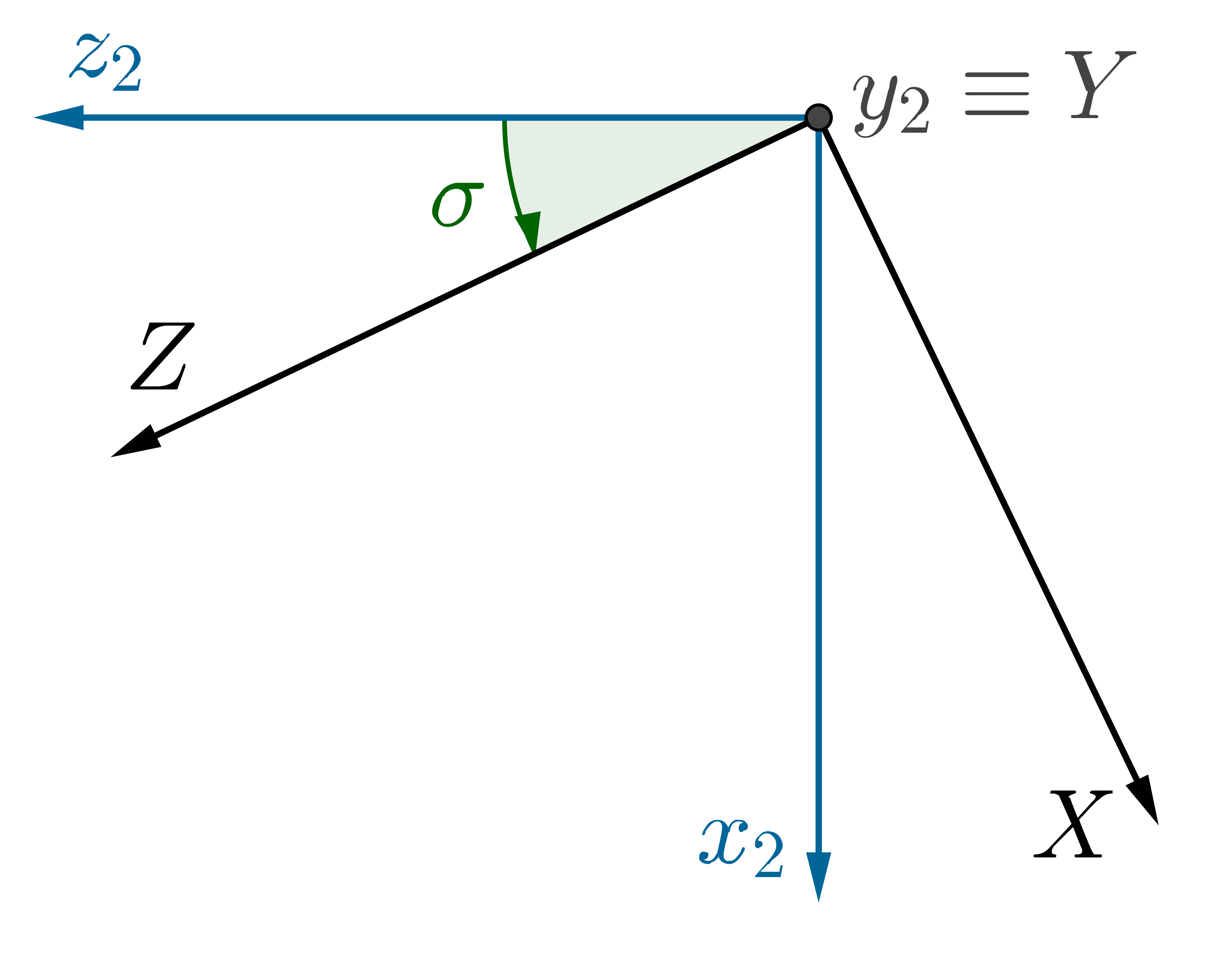}
    \caption{Euler angles ($\psi$ not shown) transforming the inertial reference frame $x,y,z$ into the spacecraft's reference frame $X,Y,Z$. Euler frequencies are also shown. On the left both $z$ and $\dot{\nu}$ are directed outwards. The same is true for $Y$ on the right. See the text for details.}
    \label{fig:Euler}
\end{figure}

To describe the motion from the external observer point of view we define a non-rotating reference frame and introduce the Euler angles (see Fig.~\ref{fig:Euler}).
During its journey in the Solar System, New Horizons must keep the main antenna towards Earth. To avoid the Euler angle singularity, when precession and spin have the same direction, we define the non-rotating reference frame $x,y,z$ such that the first axis $x$ is directed away from the Earth. Therefore, in the unperturbed spin mode, the antenna is directed towards the negative $x$ axis of the inertial reference frame. The Euler angles are successively constructed as a rotation around the $y$ axis, defining the precession $\dot{\psi}$, followed by rotation around a new $z_1$ axis, defining the nutation $\dot{\nu}$, and finally followed by a rotation around the new $y_2=Y$ axis, defining the spin $\dot{\sigma}$. Without loss of generality, we can set the Euler angle initial conditions to
\begin{equation}
\label{eq:Euler_ang_ini}
\psi(0)=0,\qquad \nu(0)=\pi/2,\qquad \sigma(0)=0.
\end{equation}
The spacecraft's angular velocity is just the sum of the Euler frequencies $\bm{\omega}= \bm{\dot{\psi}} + \bm{\dot{\nu}} + \bm{\dot{\sigma}}$. Using Eq.~\eqref{eq:Euler_eq_sol}, noting that if the effect is expected to be small then $\psi,\dot{\psi},\dot{\nu}$, but not $\sigma,\dot{\sigma}$, must be small, and $\nu\simeq\pi/2$, we can define $\epsilon\ll 1:\;\nu\equiv\frac{\pi}{2}-\epsilon\;\Rightarrow\;\sin\nu\simeq1,\ \cos\nu\simeq\epsilon$. Consequently, the angular velocity can be approximated by
\begin{subequations}
    \label{eq:w_approx}
    \begin{align}
    \label{eq:w_approxa} \delta\omega_X &\simeq \dot{\psi}\cos\sigma-\dot{\nu}\sin\sigma, \\
    \label{eq:w_approxb} \omega_0       &\simeq \dot{\psi}\epsilon+\dot{\sigma}, \\
    \label{eq:w_approxc} \delta\omega_Z &\simeq \dot{\psi}\sin\sigma+\dot{\nu}\cos\sigma.
    \end{align}
\end{subequations}
The solution of Eq.~\eqref{eq:w_approxb}~can be immediately obtained,
\begin{equation}
\label{eq:dot_sigma}
\dot{\sigma}\simeq\omega_0\quad \Rightarrow\quad \sigma(t)\simeq\omega_0 t,
\end{equation}
by noting that both $\epsilon,\dot{\psi}$ are small and can be neglected, and using the initial conditions from Eq.~\eqref{eq:Euler_ang_ini}. Knowing $\sigma$, the remaining Eqs.~\eqref{eq:w_approxa}~and~\eqref{eq:w_approxc}~can now be inverted to obtain $\dot{\psi},\dot{\nu}$, and integrated to retrieve the Euler angles $\psi(t),\nu(t)$, using Eqs.~\eqref{eq:Euler_eq_sol}~and~\eqref{eq:Euler_ang_ini}.

The solution, in the case $\Omega\neq\omega_0$, is
\begin{subequations}
    \label{eq:psi_nu}
    \begin{align}
    \label{eq:psi_nua} \psi(t) &=
    \frac{
        (\alpha_1\Omega-\alpha_2\omega_0)\sin\Omega t\cos\omega_0t
    }{\omega_0^2-\Omega^2} \nonumber \\
    &+\frac{
        \left[	
        \left(\frac{\alpha_1}{\omega_0}\right)(\omega_0^2-\Omega^2)
        +(\alpha_2\Omega-\alpha_1\omega_0)\cos\Omega t
        \right] \sin\omega_o t
    }{\omega_0^2-\Omega^2}, \\
    \label{eq:psi_nub} \nu(t) &=
    \frac{
        \left(\frac{\Omega}{\omega_0}\right)(\alpha_1\Omega-\alpha_2\omega_0)
    }{\omega_0^2-\Omega^2}
    -
    \frac{
        (\alpha_1\Omega-\alpha_2\omega_0)\sin\Omega t\sin\omega_0 t
    }{\omega_0^2-\Omega^2}
    \nonumber \\
    &+\frac{
        \left[
        \left(\frac{\alpha_1}{\omega_0}\right)(\omega_0^2-\Omega^2)
        +(\alpha_2\Omega-\alpha_1\omega_0)\cos\Omega t
        \right] \cos\omega_0t
    }{\omega_0^2-\Omega^2}.
    \end{align}
\end{subequations}
The condition $\Omega=\omega_0$ is equivalent to
\begin{equation}
\sqrt{(B/A-1)(B/C-1)}=1 \;\Leftrightarrow\; B=A+C
\end{equation}
which implies that $R=\frac{A}{C}$\ and $\alpha_1=\alpha_2=\frac{M}{C\omega_0}$. Therefore, the solution in this case becomes
\begin{subequations}
    \label{eq:psi_nu_zero}
    \begin{align}
    \label{eq:psi_nu_zeroa} \psi(t) &=
    \frac{\alpha_1}{2\omega_0}(2\sin\omega_0t-\sin2\omega_0 t) \\
    \label{eq:psi_nu_zerob} \nu(t) &=
    \frac{\alpha_1}{\omega_0}(\sin^2\omega_0t-1+\cos\omega_0t),
    \end{align}
\end{subequations}
where there is not a resonance effect, as it should be expected, since this is just the gyroscopic stabilisation effect that does not have anything resembling resonance. Equations~\eqref{eq:psi_nu}~or~\eqref{eq:psi_nu_zero}~(depending on whether $B\neq A+C$ or $B=A+C$), together with Eq.~\eqref{eq:dot_sigma}, are the general solution for the passive spin-stabilised mode of the spacecraft, with the small applied moment of Eq.~\eqref{eq:moment_general}. In the case of small $M$ the Euler angles $\psi,\nu$ remain small, confirming the validity of the approximate solution.

\subsubsection{Three-Axis Mode Case}
\label{subsec:threeaxis_case}

In the case of the three-axis mode, in the absence of thrust, and since the thermal moment has the direction of a principal axis, the effect on rotation is simply a small one around the $Z$ axis, determined by
\begin{equation}
\label{eq:3_axis_mode}
\dot{\delta\omega_Z} = M/C.
\end{equation}

\newpage
%


\end{document}